\documentclass[a4paper,11pt]{article}
\usepackage{jheppub} % for details on the use of the package, please see the JINST-author-manual
\usepackage{lineno}
\usepackage{multirow}
\usepackage{graphicx}
\usepackage{tabularx}
\usepackage{mathtools}
\usepackage{slashed}
\usepackage{booktabs}
\usepackage{url}
\usepackage[maxfloats=100]{morefloats}
\usepackage{color,xcolor}
\usepackage[normalem]{ulem}
\usepackage{subfig}
\newcommand{\MGMCatNLO}{MadGraph5\_aMC@NLO} 
\newcommand{\pythia}{{\sc Pythia}}
\newcommand{\delphes}{{\sc Delphes}}
\newcommand{\fastjet}{{\sc FastJet}}
\newcommand{\met}{\ensuremath{\slashed{E}_T}}

\newcommand{\abinv}{\mbox{\ensuremath{~\mathrm{ab^{-1}}}}}
\newcommand{\PZ}{\ensuremath{\mathrm{Z}}}

\newcommand{\PW}{\ensuremath{\mathrm{W}}}

\newcommand{\kt}{\ensuremath{k_\mathrm{T}}}
\newcommand{\pt}{\ensuremath{p_\mathrm{T}}}

\newcommand{\ptll}{\ensuremath{p_{\mathrm{T},\ell\ell}}}

\newcommand{\PTeV}{\ensuremath{\mathrm{TeV}}}
\newcommand{\PGeV}{\ensuremath{\mathrm{GeV}}}

\title{Searching for Majorana Neutrinos at a Same-Sign Muon Collider}
\author[a,b]{Ruobing Jiang,}
\author[a,1]{Tianyi Yang\note{Corresponding author.},}
\author[a]{Sitian Qian,}
\author[a]{Yong Ban,}
\author[c]{Jingshu Li,}
\author[c]{Zhengyun You,}
\author[a]{Qiang Li}
\affiliation[a]{Department of Physics and State Key Laboratory of Nuclear Physics and Technology, Peking University, Beijing, 100871, China}
\affiliation[b]{School of Physical Science and Technology, Southwest University}
\affiliation[c]{
School of Physics, Sun Yat-Sen University, Guangzhou 510275, China}

\date{\today}

\emailAdd{tyyang99@pku.edu.cn}

\abstract{Majorana properties of neutrinos have long been a focus in the pursuit of possible new physics beyond the standard model, which has motivated lots of dedicated theoretical and experimental studies. A future same-sign muon collider is an ideal platform to search for Majorana neutrinos through the Lepton Number Violation process: $\mu^{+}\mu^{+} \rightarrow \PW^{+}\PW^{+}$. Specifically, this t-channel kind of process is less kinematically suppressed and has a good advantage in probing Majorana neutrinos at high mass regions up to 10 \PTeV. In this paper, we perform a detailed fast Monte Carlo simulation study through examining three different final states: 1) pure-leptonic state with electrons or muons, 2) semi-leptonic state, and 3) pure-hadronic state in the resolved or merged categories. Furthermore, we perform a full simulation study on the pure-leptonic final state to validate our fast simulation results.}

\begin{document}

\maketitle

\section{Introduction}

\label{sec:intro}

The discovery of the Higgs boson in 2012~\cite{plb:2012gu,plb:2012gk} marked a triumph of the Standard Model (SM) of particle physics and the Large Hadron Collider (LHC). The LHC and the High-Luminosity LHC (HL-LHC), together with other future colliders such as muon colliders in design, will further explore the SM and search for physics beyond that. However, despite the enormous success of SM and LHC, the observation of neutrino oscillations has confirmed that at least two SM neutrinos have small but nonzero masses, and there is flavor mixing among three-generation light neutrinos. This provides a compelling hint of physics beyond the SM, because the right-handed components of neutrinos and the tiny neutrino masses as well as the flavor mixtures of the lepton sectors are not expected in the original SM~\cite{Yang:2023ojm}. Therefore, searching for Majorana neutrinos in order to explain the origin of neutrino masses is extremely important. 

Majorana neutrinos have been studied at various types of colliders, including the LHC~\cite{CMS:2018iaf,ATLAS:2019kpx,CMS:2019lwf,CMS:2018jxx,ATLAS:2018dcj,CMS:2022rqc}, electron-positron collider~\cite{DELPHI:1996qcc,Almeida:2001kq,Zhang:2018rtr}, electron-electron collider~\cite{Wang:2016eln}, electron-proton collider~\cite{Gu:2022muc} and muon-muon collider~\cite{Kwok:2023dck,Li:2023tbx}. In this paper, we focus on the inverse $0\nu\beta\beta$-like channel through same-sign muon collisions~\cite{Yang:2023ojm}: $\mu^+\mu^+\rightarrow \PW^{+}\PW^{+}$, and perform a research on the relationship between the mass of Majorana neutrinos and the square of mixing element. The constraints on the squared mixing element between the muon and the Majorana neutrino are derived in the Majorana neutrino mass range of 100\PGeV -- 40\PTeV. The results show that our signal process have unique advantages when the mass is greater than 10\PTeV.

\section{Majorana Neutrinos and Type-I Seesaw Model}
\label{sec:Majorana}
Majorana neutrinos are a common feature of many extensions of the SM, motivated by their roles in explaining the generation of neutrino masses. The simplest renormalizable extension of the SM for understanding the smallness of the left-handed neutrino masses is defined by the interaction Lagrangian :
\begin{equation}
     -L_{y}=y_{\ell\alpha}\overline{L_{\ell}}\widetilde{\Phi}\ N_{R\alpha}+\mathrm{H.c}.
\end{equation}
 where $y_{\ell\alpha}$ is the dimensionless complex Yukawa couplings, $\alpha =1, 2, ...., N$ is the singlet neutral fermions flavor index, $\overline{L_{\ell}}$ corresponds to the lepton field, $\widetilde{\Phi}$ corresponds to higgs field, $N_{R\alpha}$ are singlet neutral fermions in SM. This Lagrangian generates a Dirac mass $M_{D}=y_{\ell\alpha}\nu$, where $\nu$ is the vacuum expectation value. Since the right-handed neutrinos carry no SM gauge charges, in order to preserve gauge invariance, we can introduce a new term of Lagrangian for Majorana mass: 
\begin{equation}
-L_{M}=\frac{1}{2}(M_{N})_{\alpha\beta}\overline{N}^{c}_{R\alpha}N_{R\beta}+\mathrm{H.c}.
\end{equation}
The two Lagrangians above together lead to the following neutrino mass matrix:
\begin{equation}\left[
    \begin{array}{cc}
       0  &  M_{D} \\
       M^{T}_{D}  & M_{N}
    \end{array}
    \right]
\end{equation}
So, the light neutrino masses and mixing element are given by the diagonalization of the effective mass matrix:
$M_{\nu}\simeq\-M_{D}M^{-1}_{N}M^{T}_{D}$, $V_{\ell N_{\alpha}}\sim\ M_{D}M^{-1}_{N}$~\cite{Deppisch:2015qwa}.
Because Majorana neutrinos can only couple to the SM through mixing with SM neutrinos, so we can establish the relationship between the SM neutrinos mass and the Majorana neutrinos mass by~\cite{CMS:2022rqc}: 
 \begin{equation}
     m_\nu=y_{\ell\alpha}^{2}\nu^{2}/m_{N}.
 \end{equation}
 
\par This is the most famous model including Majorana neutrinos: Type-I Seesaw model. This mechanism can explain neutrinos' small masses and the variations of neutrinos in SM. It directly shows that the smallness of SM neutrinos masses can be explained by a suppression due to the high mass of new particles in the Type-I seesaw mechanism.
 \par From the above discussion, the massive Majorana neutrinos can testify the small mass of SM neutrinos and heavy Majorana neutrinos are mixed with SM neutrinos, which is characterized by the mixing element $|V_{\ell N_{\alpha}}|^{2}$, between an SM neutrino in its left-handed interaction state and a heavy Majorana neutrino in its mass eigenstate~\cite{CMS:2022rqc}. Therefore, it is clear that there are two key aspects of the Type-I seesaw mechanism that can be probed experimentally, the Majorana neutrino mass $M_{N}$, and the mixing element $|V_{\ell N_{\alpha}}|$. And Majorana neutrino mass will put a constraint on $|V_{\ell N_{\alpha}}|^{2}$. For the Majorana neutrinos masses above the EW boson masses, the highest sensitivity is expected for the heavy Majorana neutrino searches in the tri-lepton or same-sign di-lepton channels. Limits on the mixing elements extend down to about $10^{-5}\sim 10^{-6}$ for neutrino masses $M_{N}$ at several GeV~\cite{CMS:2022rqc}.

\section{Majorana Neutrinos Study at Same-Sign Muon Collider}
\label{sec:muon}

A muon--muon collider with the center-of-mass (c.m.) energy at the multi-\PTeV\ scale has received much-revived interest~\cite{Daniel20} recently, which has several advantages compared with both hadron--hadron and electron--electron colliders~\cite{Mario16,Dario18,Hind21,Mauro20}. On the one hand, as massive muons emit much less synchrotron radiation than electrons, muons can be accelerated in a circular collider to higher energies with a much smaller circumference. On the other hand, because the proton is a composite particle, muon--muon collisions are cleaner than proton--proton collisions and thus can lead to higher effective c.m. energy. However, due to the short lifetime of the muon, the beam-induced background (BIB) from muon decays needs to be examined and constrained properly. Based on a realistic simulation at $\sqrt{s}=1.5$\PTeV\ with BIB included~\cite{Nazar20}, found that the coupling between the Higgs boson and the b-quark can be measured at the percentage level with order\abinv\ of collected data.

%At the \PTeV \ scale muon collider, searching for Majorana neutrinos can be an interesting topic. Previous detailed studies through the leptonic di-flavor and di-number violation processes at a high energy same-sign muon collider has already been performed to search for Majorana neutrinos~\cite{Yang:2023ojm}. One of the most recent study~\cite{CMS:2022rqc} shows that the signal cross section could reach $\mathrm{fb}$ level when there’re three generations of heavy Majorana neutrinos ($N_{I}, I=1,2,3$) with their hierarchical masses satisfy $M_2\ll M_1\ll M_3$. 

\begin{figure}
\centering
\subfloat[\label{fig:a}]{
\includegraphics[width=7cm]{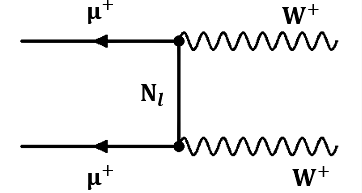}}
\subfloat[\label{fig:b}]{
\includegraphics[width=7cm]{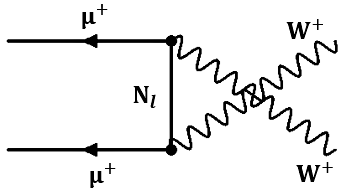}}
\caption{Signal processes at a same sign muon collider :(a) t-channel, (b) u-channel}
\label{fig:figure1}
 \end{figure}

In this study, we perform a search of Majorana neutrinos at a same-sign muon collider. We target two benchmark scenarios in this study, i.e., a c.m. energy of 1\PTeV \ and 10\PTeV\ with a luminosity of 1\abinv \ , and we add a luminosity of 10\abinv\ for 10\PTeV\ channel to get the most sensitive result. The processes related to the mediation by Majorana neutrinos at a same-sign muon collider are shown in Fig.~\ref{fig:figure1}, which is sensitive to the \PTeV-seesaw scenario. Based on the decay channels of two $\PW^{+}$ bosons, we can divide our final states into three channels: a pure-leptonic channel with two leptons, a semi-leptonic channel with one lepton and two jets, and a pure-hadronic channel with four jets. Furthermore, we also analyze the impact of fatjets at several TeV c.m. energy.  

In the following sections, we firstly discuss the kinematic properties of $\mu^{+}\mu^{+}\rightarrow \PW^{+}\PW^{+}$ process and introduce our cut-flow strategy to reduce the background events. Then, we present our numerical analysis results and discuss the detection possibilities in all three final states. To compare the consistency of fast simulation and full simulation results, the pure-leptonic final state is studied by full simulation and compared with the fast simulation. Lastly, we give the results of limits and compare them with previous analyses.
 
\section{Simulation and Analysis Framework}
\label{sec:frame}
Both signal and background events are simulated with \MGMCatNLO~\cite{Buckley:2011ms}, then showered and hadronized by \pythia8~\cite{Sjostrand:2014zea}. The final state jets are clustered using \fastjet~\cite{Cacciari:2011ma} with the \kt~\cite{Cacciari:2008gp} algorithm at a fixed cone size of $R_{\rm jet}=0.5$. We used \delphes~\cite{deFavereau:2013fsa} version 3.0 to simulate detector effects with the default card for the muon collider detector~\cite{mucard}. Note that the present jet tagging techniques for muon colliders are in a preliminary stage~\cite{Nazar20} and have a large potential to be improved.
In this section, we discuss in detail the three final states of the signal process introduced above: pure-leptonic state, semi-leptonic state, and hadronic state. There are seven corresponding backgrounds:
\begin{itemize}
    \item $\mu^{+}\mu^{+}\rightarrow \PW^{+}\PW^{+}\Tilde{\nu_{\mu}}\Tilde{\nu_{\mu}}$,
    \item $\mu^{+}\mu^{+}\rightarrow \PZ\PW^{+}\mu^{+}\Tilde{\nu_{\mu}}$,
    \item $\mu^{+}\mu^{+}\rightarrow \PW^{+}\mu^{+}\nu_{\mu}\nu_{\mu}$,
    \item $\mu^{+}\mu^{+}\rightarrow \PZ\mu^{+}\mu^{+}$,
    \item $\mu^{+}\mu^{+}\rightarrow \PZ\PZ\mu^{+}\mu^{+}$,
    \item $\mu^{+}\mu^{+}\rightarrow \PW^{+}\PW^{-}\mu^{+}\mu^{+}$,
    \item $\gamma\gamma\rightarrow \PW^{+}\PW^{-}.$
\end{itemize}

\subsection{Pure-leptonic channel}

\begin{figure}
\centering
\subfloat[\label{fig:a}]{
\includegraphics[width=7cm,height=4cm]{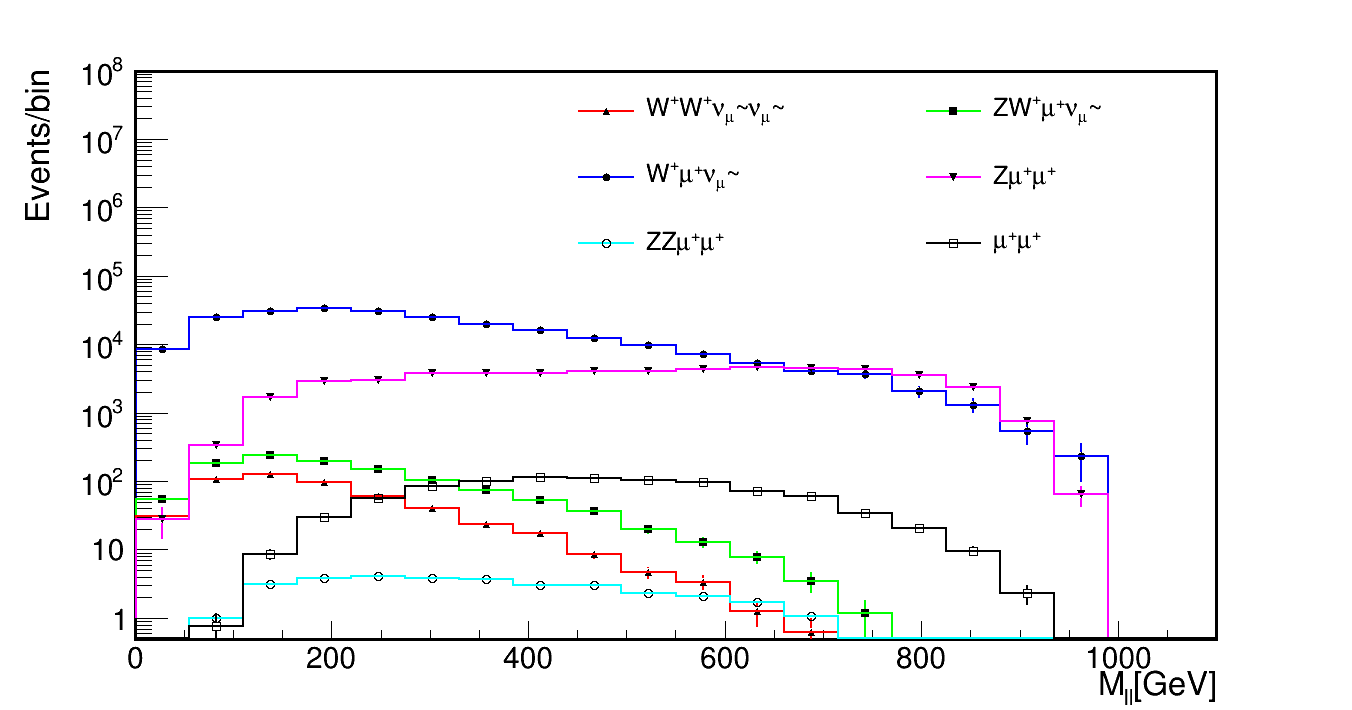}}
\subfloat[\label{fig:b}]{
\includegraphics[width=7cm,height=4cm]{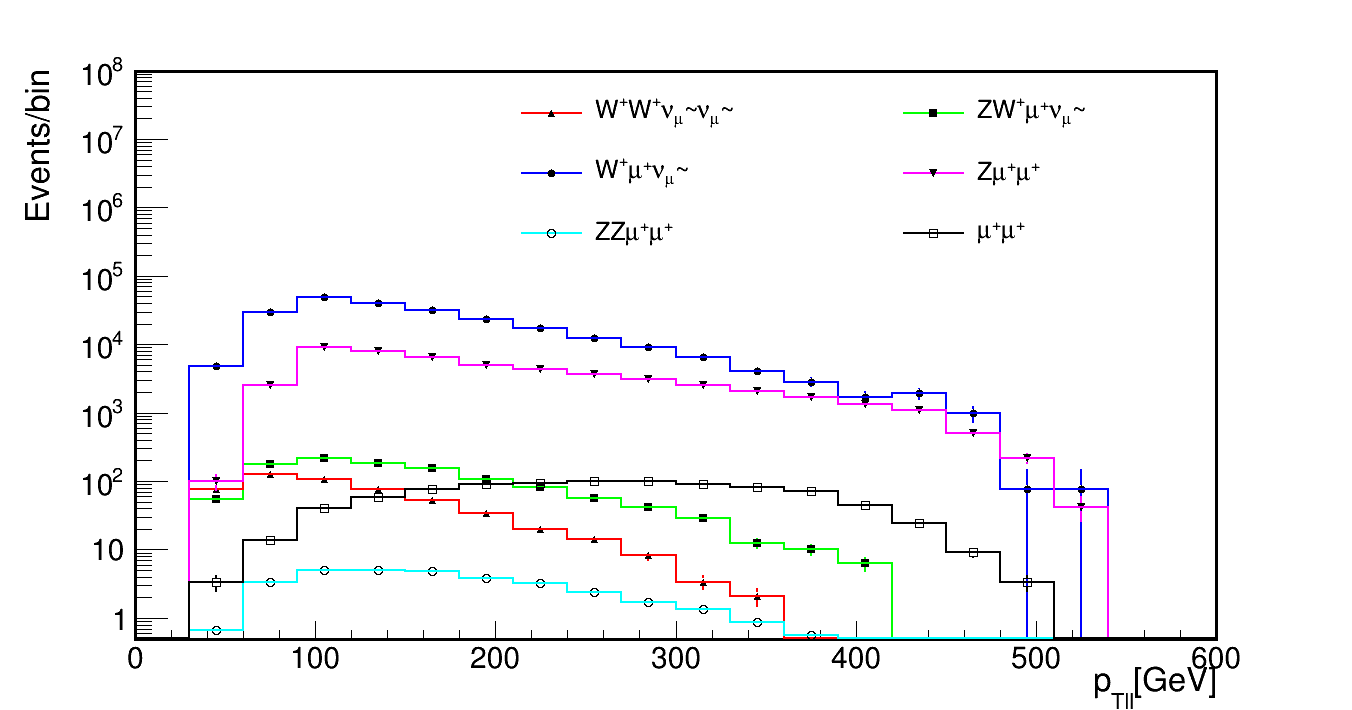}}
\\
\subfloat[\label{fig:c}]{
\includegraphics[width=7cm,height=4cm]{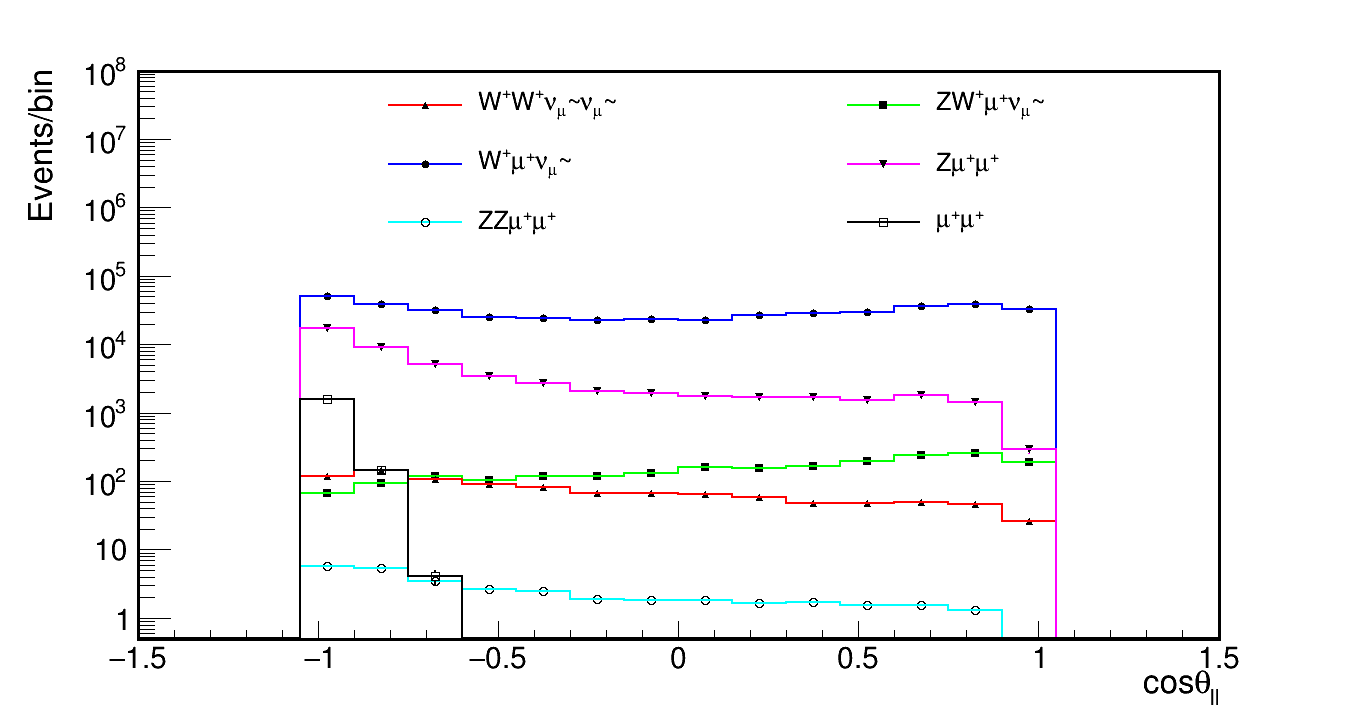}}
\subfloat[\label{fig:d}]{
\includegraphics[width=7cm,height=4cm]{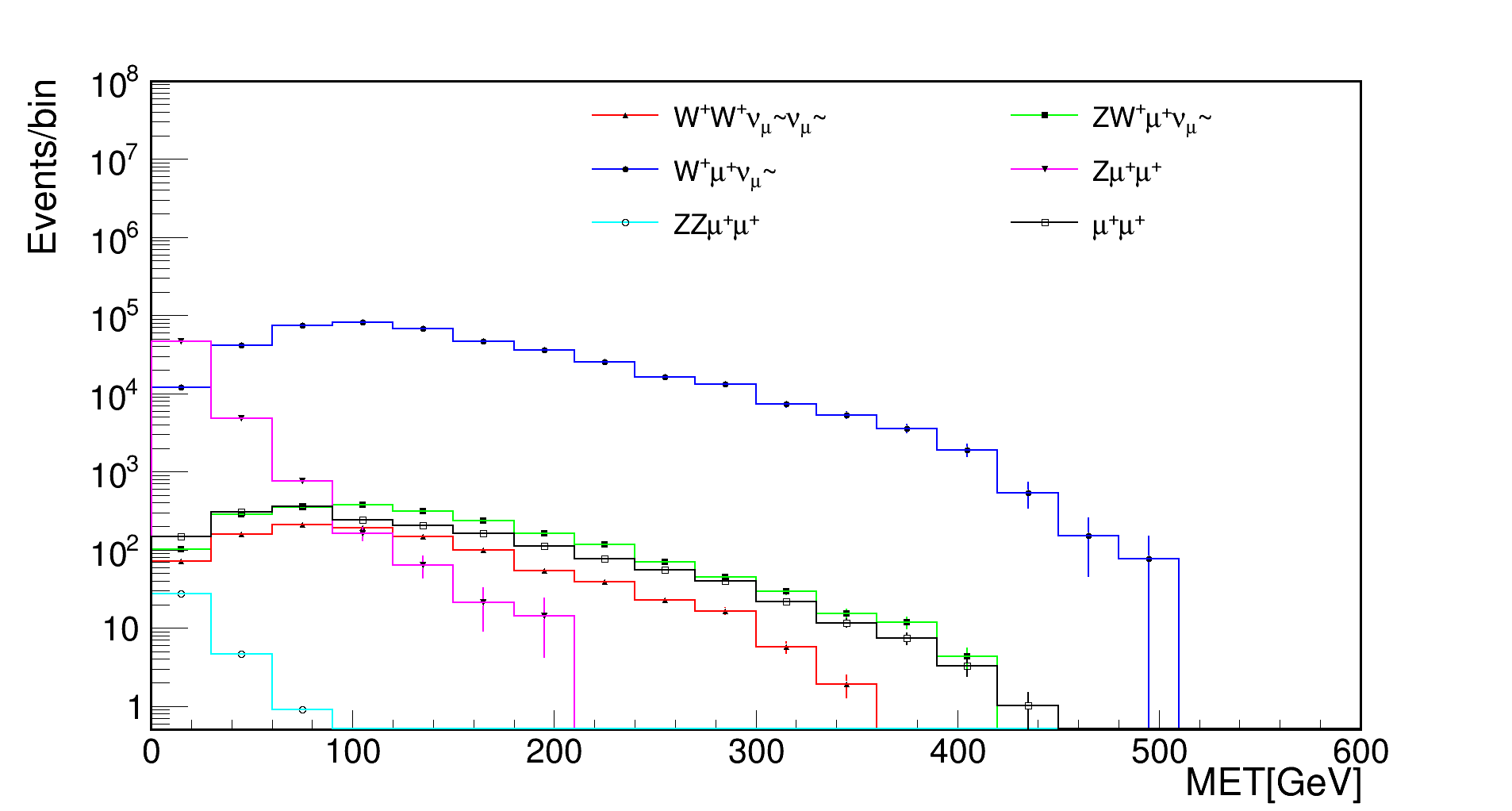}}
\caption{Simulation results of pure-leptonic channel, $\sqrt{s}=1\PTeV, \mathcal{L}=1\abinv$.
(a) invariant mass $M_{\ell\ell}$ distribution, (b) transverse momentum of leading lepton $\ptll$ distribution, (c) $\cos{\theta_{\ell\ell}}$ ($\theta_{\ell \ell}$ is the angle between two leptons in final states) distribution, and (d) missing transverse energy $\met$ distribution. }
\label{fig:figure2}
 \end{figure}

\begin{figure}
\centering
\subfloat[\label{fig:a}]{
\includegraphics[width=7cm,height=4cm]{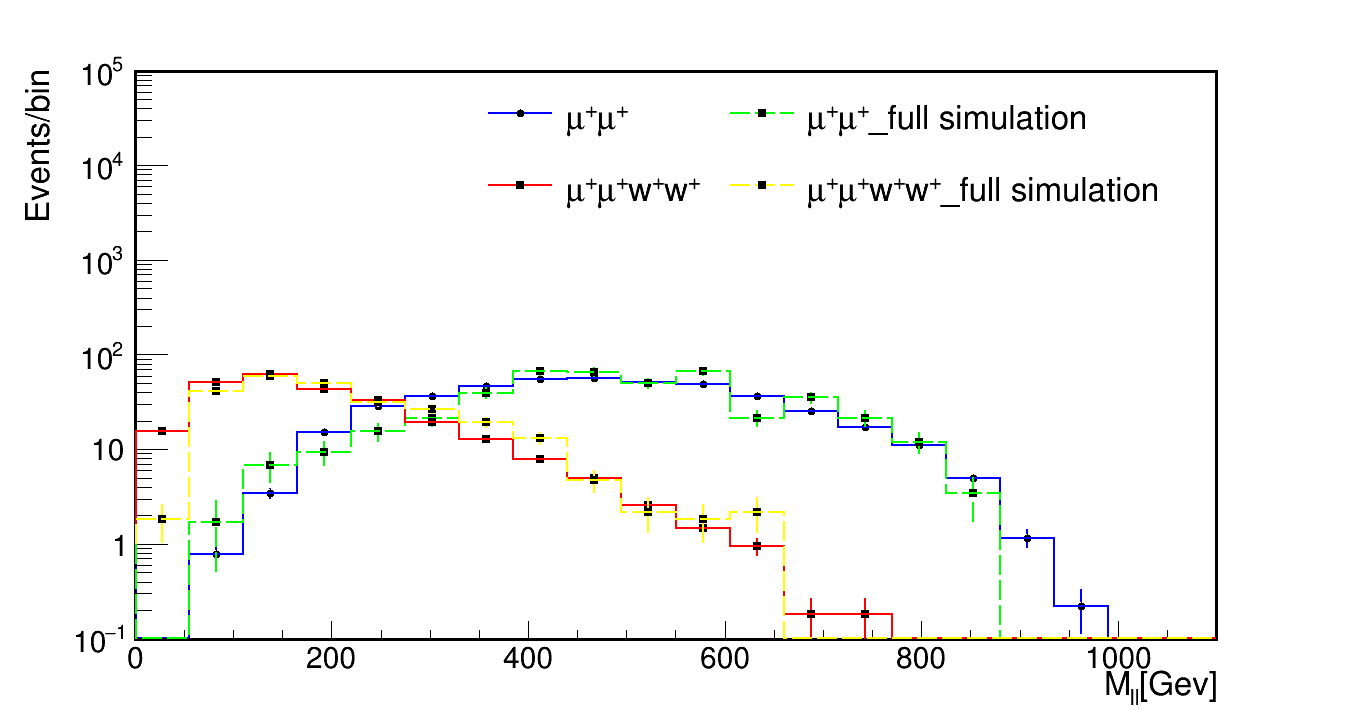}}
\subfloat[\label{fig:b}]{
\includegraphics[width=7cm,height=4cm]{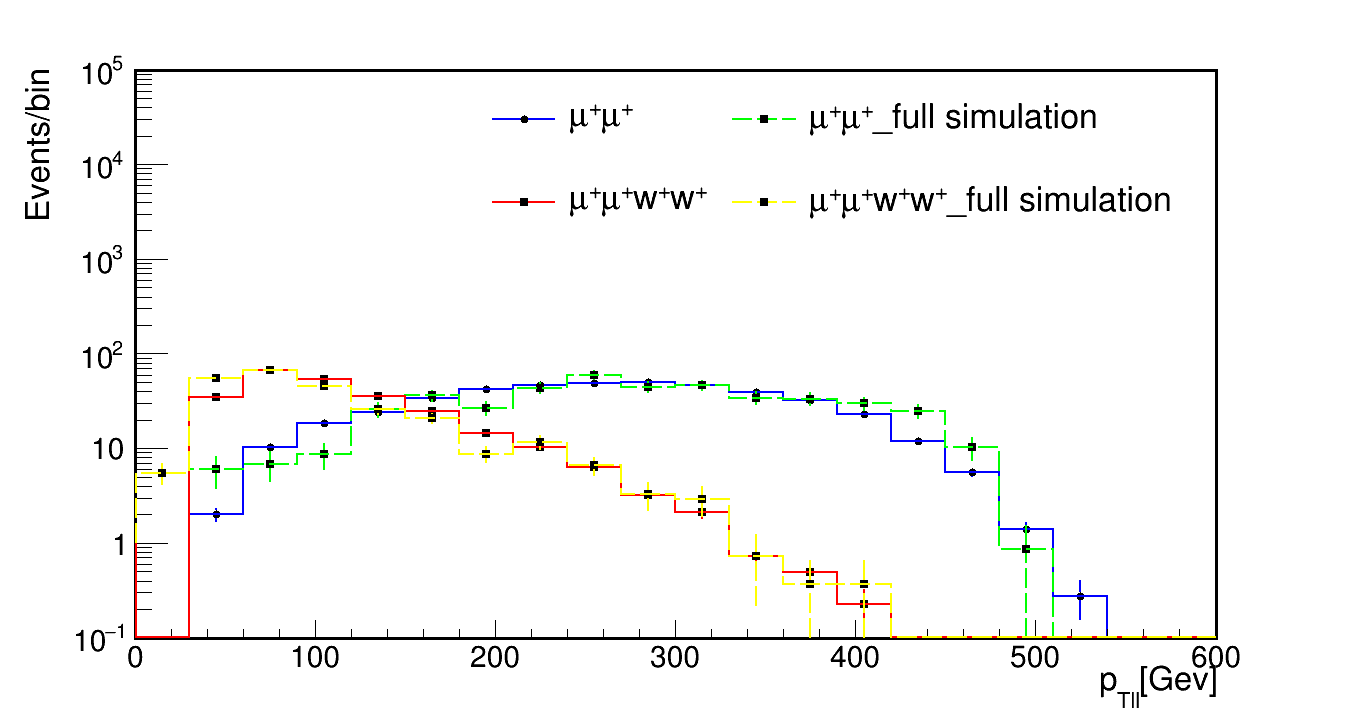}}
\\
\subfloat[\label{fig:c}]{
\includegraphics[width=7cm,height=4cm]{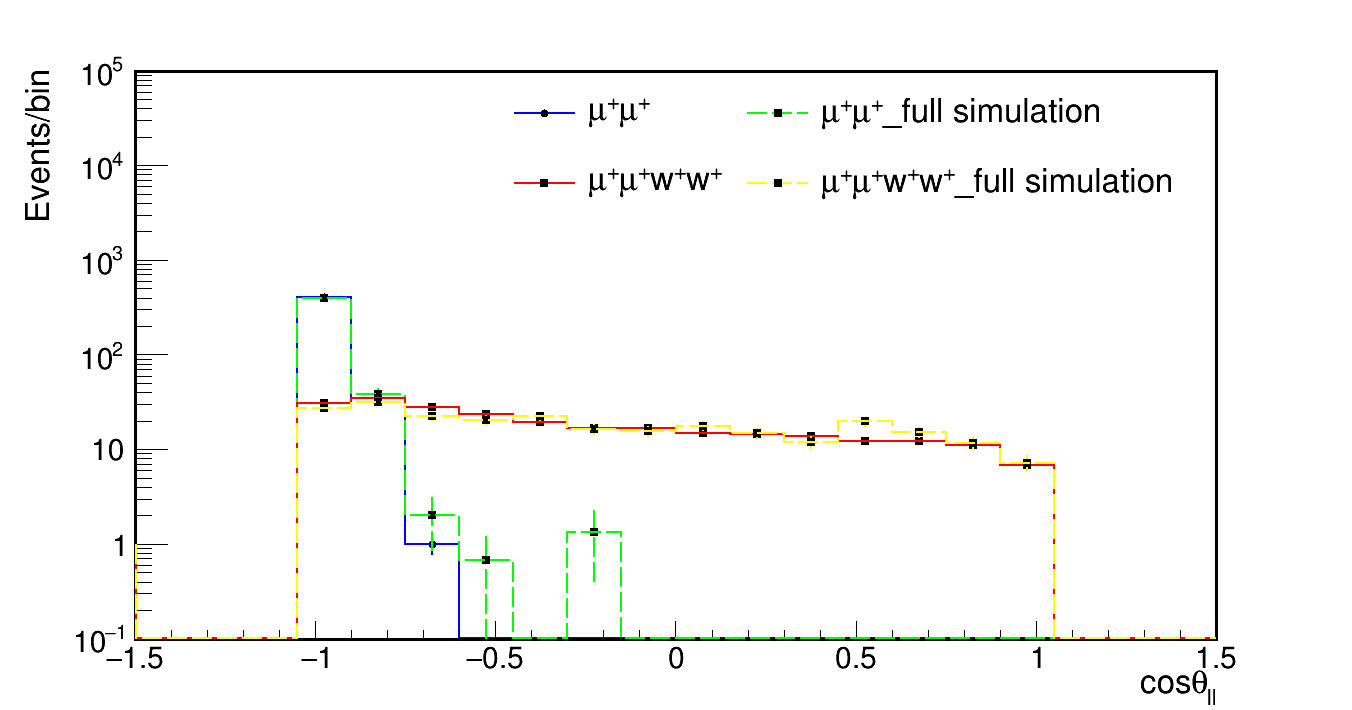}}
\subfloat[\label{fig:d}]{
\includegraphics[width=7cm,height=4cm]{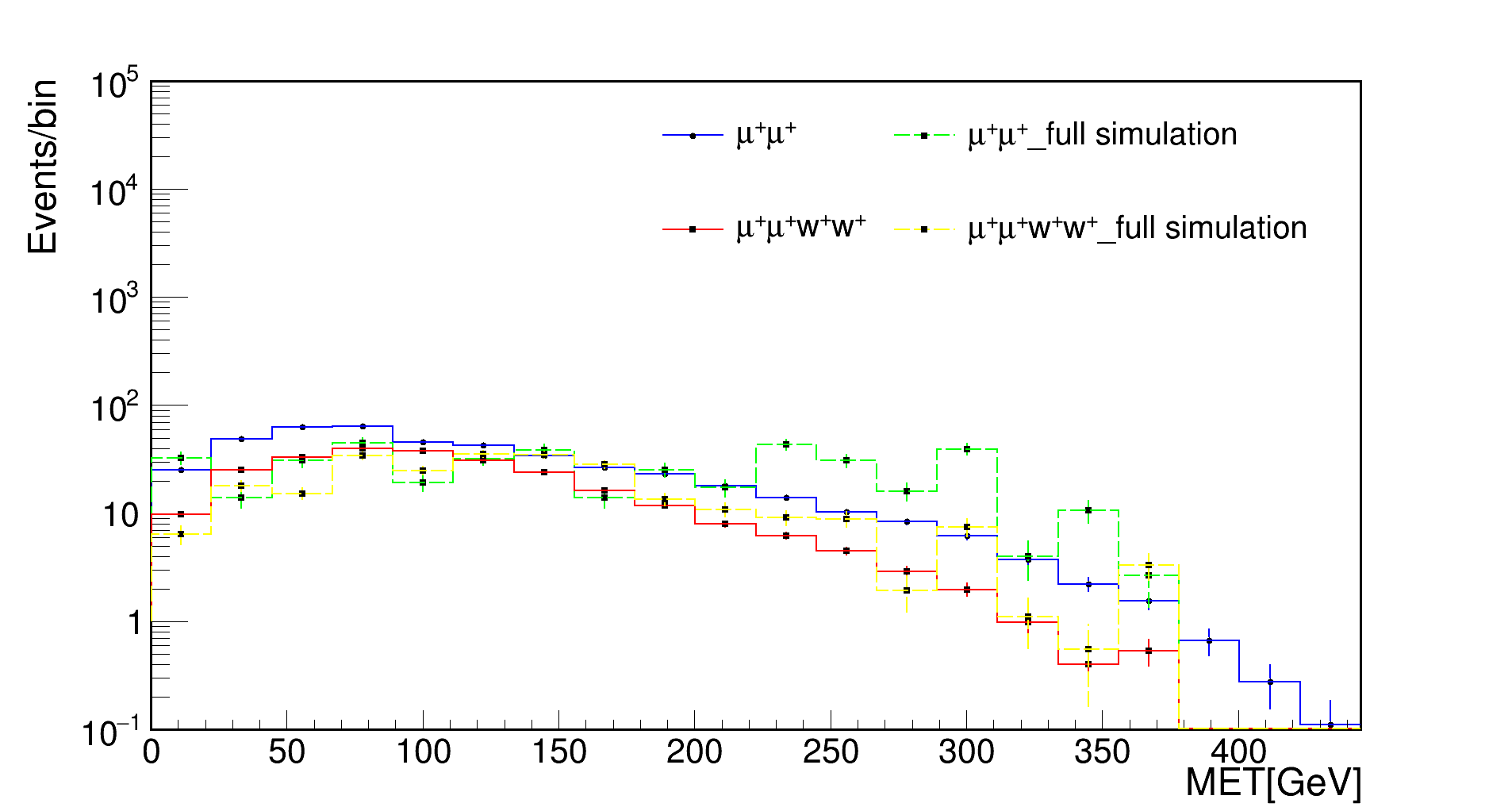}}
\caption{Comparison of fast simulation and full simulation, $\sqrt{s}=1\PTeV, \mathcal{L}=1\abinv$.
(a) invariant mass $M_{\ell\ell}$ distribution, (b) transverse momentum of leading lepton $\ptll$ distribution, (c) $\mathrm{\cos{\theta_{\ell\ell}}}$ ($\theta_{\ell \ell}$ is the angle between two leptons in final states) distribution, and (d) missing transverse energy $\met$ distribution. }
\label{fig:figure3}
 \end{figure}

 We apply constraints on the channel $\mu^{+}\mu^{+}\rightarrow \PW^{+}\PW^{+}\rightarrow\ 2\ell+\met$ at $\sqrt{s}=1\PTeV$ and $\mathcal{L}=1\abinv$ are: the events must include exactly two leptons with transverse momentum $p_\mathrm{T}>20\PGeV$, absolute pseudo-rapidity $|\eta_{\ell}|\ <2.5$, and $\Delta R_{\ell \ell}\ >0.4$, where $\Delta R=\sqrt{(\Delta\phi)^{2}+(\Delta\eta)^{2}}.$ Fig.~\ref{fig:figure2} shows some typical distributions with fast simulation, including invariant mass $M_{\ell \ell}$, transverse momentum of the leading lepton $\ptll$, $\cos{\theta_{\ell \ell}}$ ($\theta_{\ell \ell}$ is the angle between two leptons in final states) and Missing transverse energy \met\ . Some full simulation results are also shown in Fig.~\ref{fig:figure3}. 

In the full simulation, we generate stable input particles through standalone software, such as \MGMCatNLO \ , then do parton shower through \pythia8. The interaction of stable particles with the detector material is simulated by GEANT4 which is closely integrated into iLCSoft framework~\cite{ILC} previously used in CLIC experiments, and now could be used in Muon Collider studies. Both the detector response and the event reconstruction are done within a single framework such as the modular Marlin framework~\cite{Marlin}. The detector geometry is defined using the DD4hep detector description toolkit, which provides a consistent interface with both GEANT4 and Marlin environments~\cite{MuonCollider:2022ded}. We list variables distribution of signal and one background process in Fig.~\ref{fig:figure3}, we can find that the fast-simulation and the full-simulation give roughly similar distribution. However, due to the full-simulation can detect $\mu^{+}$ in final states more accurately, there is also some difference in variables distribution between full-simulation and fast simulation, especially, the distribution of \met \ , because there is not easy for full-simulation to contain all objects in final states. Among all variables, the $cos{\theta_{\ell\ell}}$ shows the most distinguishable behavior between the signals and backgrounds. The cut conditions in the pure-leptonic channel are listed in Table~\ref{tab:table1}. The significance is defined as: $S={s}/\sqrt{b} =1.96$ (CL=95\%), where $s$ and $b$ represent the number of signal and background events after all cuts, respectively. Since significance $S$ is proportion to $|V_{\ell N_{\alpha}}|^{4}$, we can obtain the limit lines for the Majorana neutrinos masses range from 100\PGeV\ to 10\PTeV.
 
 \begin{table}[h]
     \centering
     \caption{The cut-flow table in the pure-leptonic channel.}
     \label{tab:table1}
     \begin{tabular}{p{2.5cm}<{\centering}p{4cm}<{\centering}}
         \hline
            variables & limits \\
         \hline
         $M_{\ell\ell}$ & $>100.0 \PGeV$  \\
         $\ptll$ & $>120.0 \PGeV$ \\
         \met & $>100.0 \PGeV$ \\
         $cos{\theta_{\ell\ell}}$ & $<-0.95$ \\
         $|\eta_{\ell}|$ & $<2.5$ \\
        \hline
     \end{tabular}
 \end{table}
 
 \subsection{Semi-leptonic channel}

\begin{figure}
\centering
\subfloat[\label{fig:a}]{
\includegraphics[width=7cm,height=4cm]{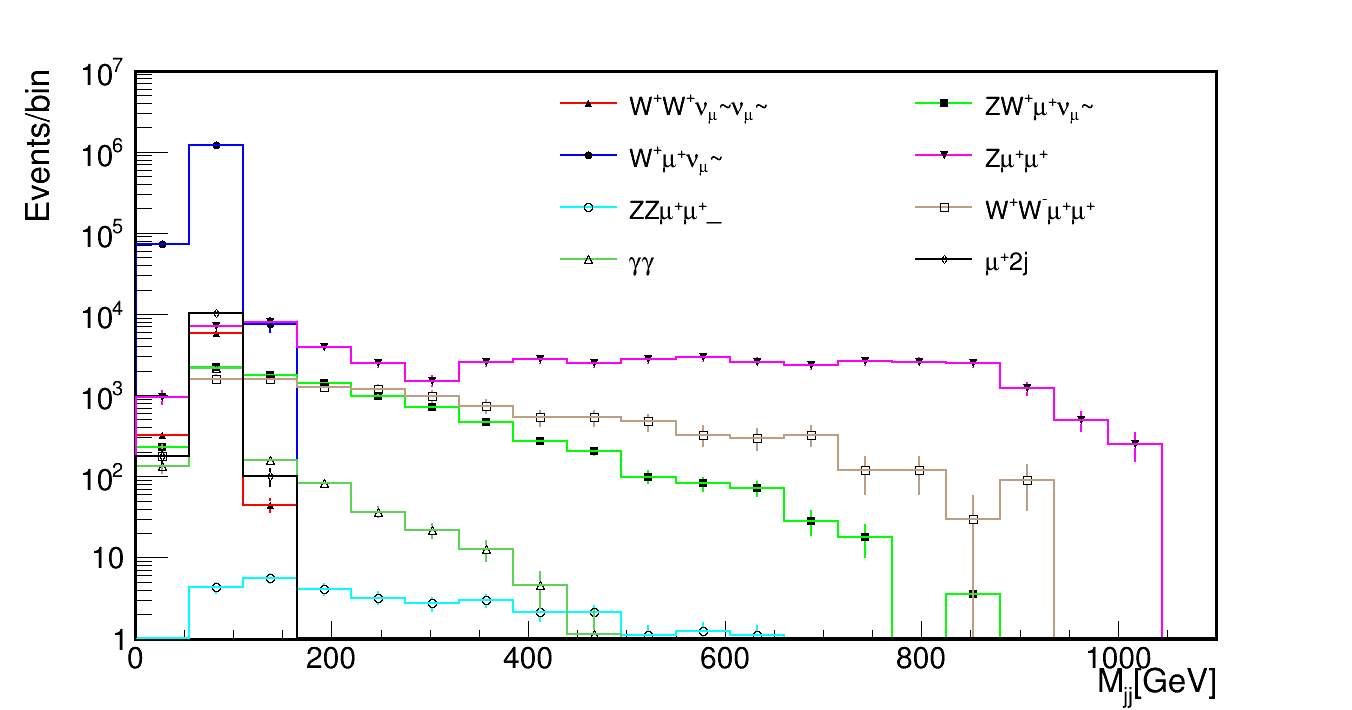}}
\subfloat[\label{fig:b}]{
\includegraphics[width=7cm,height=4cm]{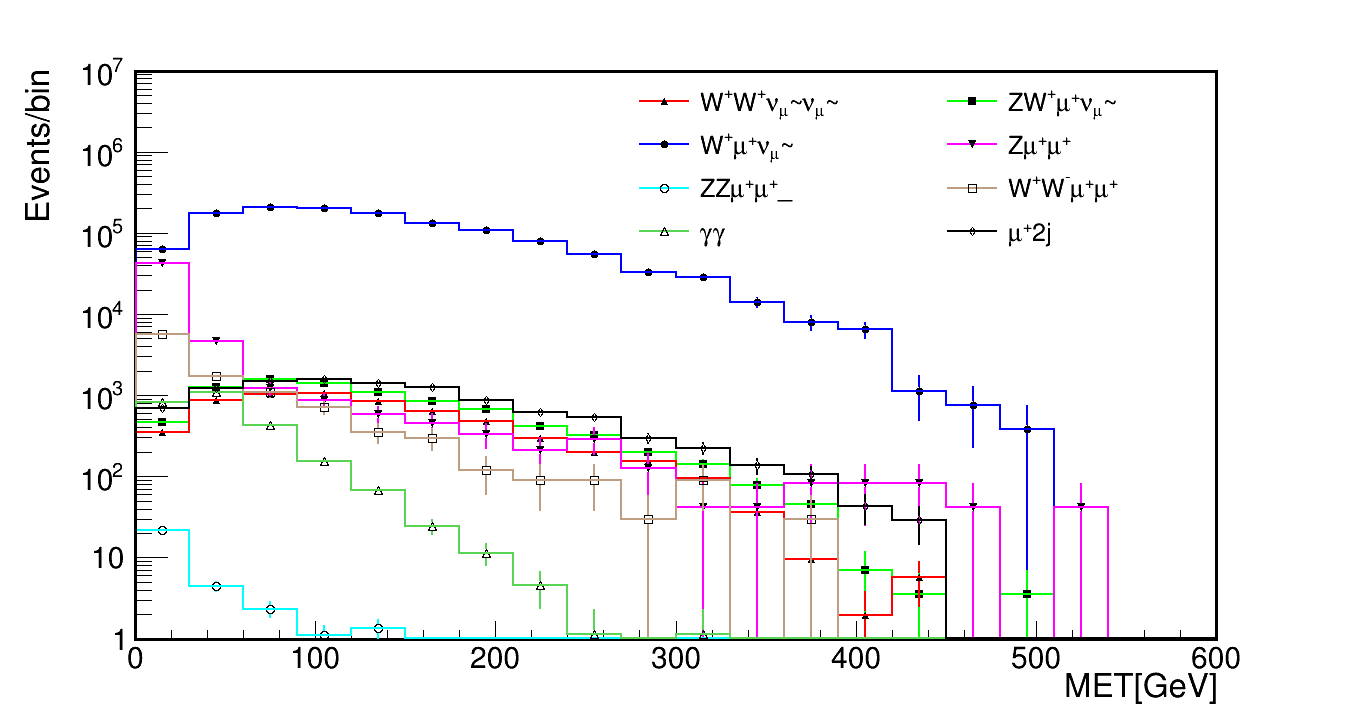}}
 \caption{Semi-leptonic channel, $\sqrt{s}=1$\PTeV, L=1\abinv. (a) the reconstructed W boson mass $M_{jj}$ distribution, and (b) $\met$ distribution.}
\label{fig:figure4}
 \end{figure}

 The constraints on the channel $\mu^{+}\mu^{+}\rightarrow \PW^{+}\PW^{+}\rightarrow\ell^{+}2j+\met$ at $\sqrt{s}=1\PTeV$ are: events must include exactly one lepton and two jets with $\pt>20\PGeV$, $|\eta_{\ell}|<2.5,\ |\eta_{j}|<4.7$, and $\Delta R_{\ell j}>0.4, \Delta R_{jj}>0.4$. Fig.~\ref{fig:figure4} shows the simulated distribution of reconstructed boson mass $M_{jj}$ and the missing transverse energy $\met\ $in the semi-leptonic process. The $\met\ $ distributions of the signal and all backgrounds are similar, so this variable can not be used to distinguish the signal from backgrounds. $\PW\ $boson can be reconstructed through $M_{jj}$ in this channel, the signal and backgrounds shows more significant difference in $M_{jj}$ distributions. The cuts applied in semi-leptonic channel are listed in Table~\ref{tab:my_table2}. The simulation results show that semi-leptonic is the least sensitive channel.
 
 \begin{table}[h]
     \centering
     \caption{The cut-flow table in the semi-leptonic channel.}
     \label{tab:my_table2}
     \begin{tabular}{p{2.5cm}<{\centering}p{4cm}<{\centering}}
     \hline
       Variables &  limits  \\
       \hline
        $M_{jj}$ & $<140.0\PGeV$\\
        $p_{\mathrm{T},jj}$ & $>50.0\PGeV$\\
        $p_{\mathrm{T},\ell}$ & $>50.0\PGeV$\\
        $|\eta_{j}|$ & $<4.7$\\
        $|\eta_{\ell}|$ & $<2.5$\\
        \hline
     \end{tabular}
 \end{table}

\subsection{Hadronic resolved channel}
 The constraints on channel $\mu^{+}\mu^{+}\rightarrow \PW^{+}\PW^{+}\rightarrow 4j$ at $\sqrt{s}=1\PTeV$ are: events must include exactly four jets or two fatjets with $\pt>20\PGeV$, $|\eta_{j}| <4.7$, and $\Delta R_{jj} >0.4.$ The four jets are classified and clustered into two reconstructed ``bosons'' $(\PW_{1},\PW_{2})$, their masses are denoted as $M_{1},M_{2}$. We use the following algorithm:
\begin{itemize}
\item Construct all possible jet pairs candidates: ($j_{1}j_{2}$, $j_{3}j_{4}$), ($j_{1}j_{3},j_{2}j_{4}$), and ($j_{1}j_{4},j_{2}j_{3}$),
\item Calculate the corresponding mass difference:
\begin{equation}
\Delta M^{2}=(M_{1}-M_{W})^{2}+(M_{2}-M_{W})^{2},  
\end{equation} 
\item Choose the minimum $\Delta M^{2}$ as the targeted jet pairs.
\end{itemize}

\begin{figure}
\centering
\subfloat[\label{fig:a}]{
\includegraphics[width=7cm,height=4cm]{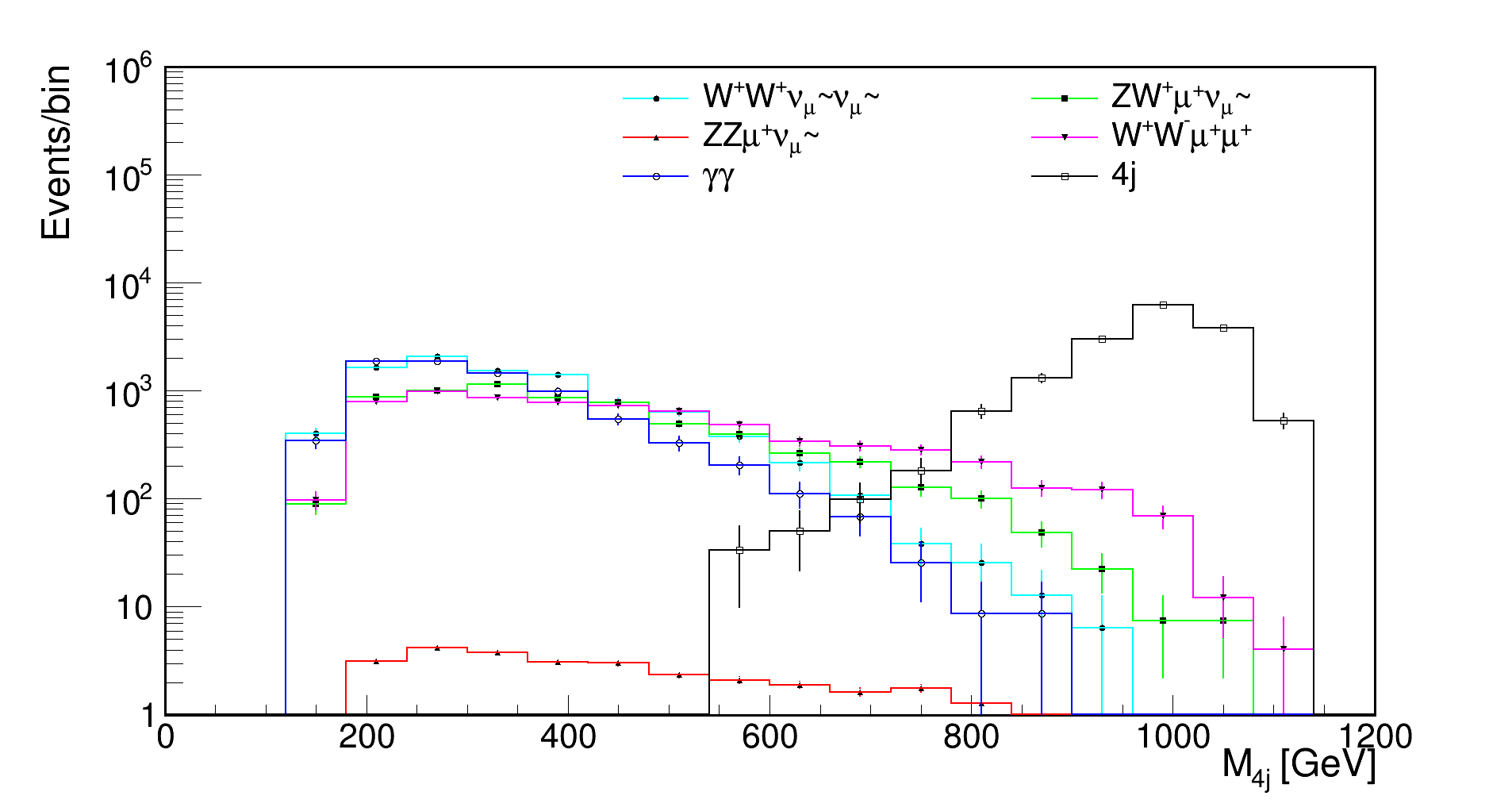}}
\subfloat[\label{fig:b}]{
\includegraphics[width=7cm,height=4cm]{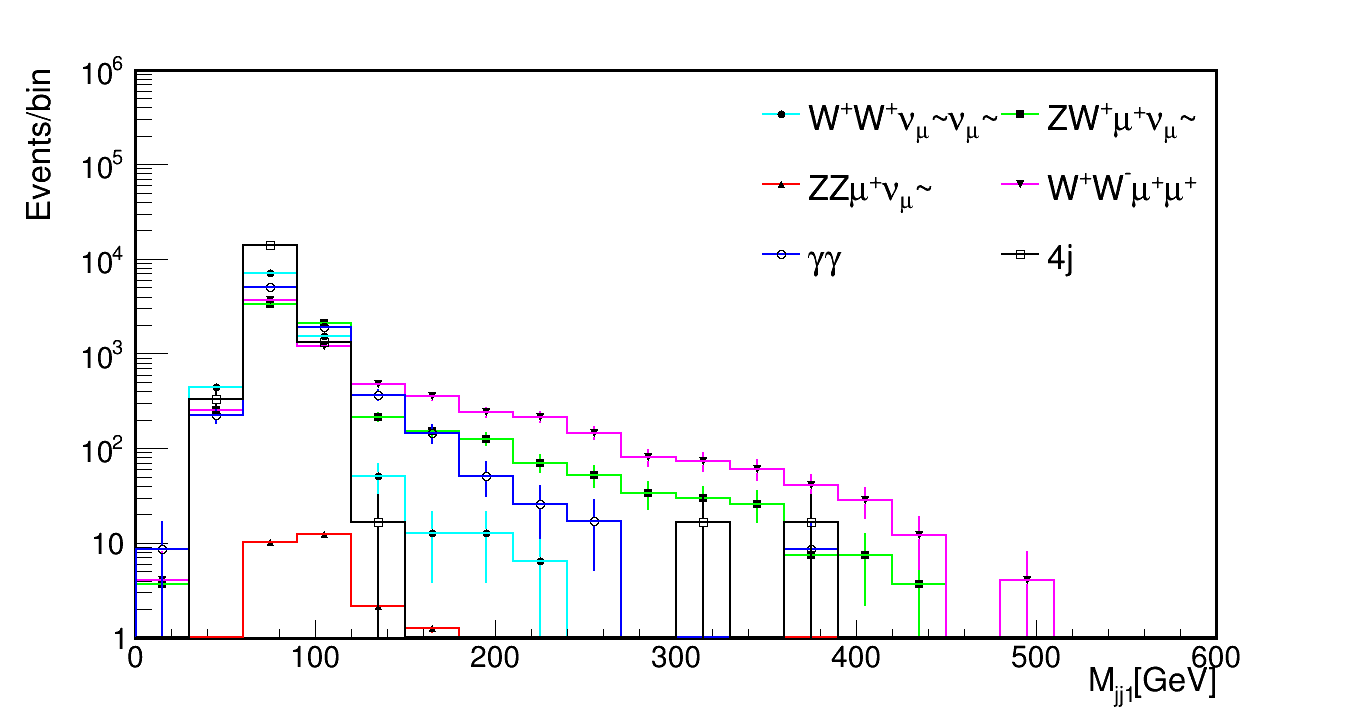}}
\\
\subfloat[\label{fig:c}]{
\includegraphics[width=7cm]{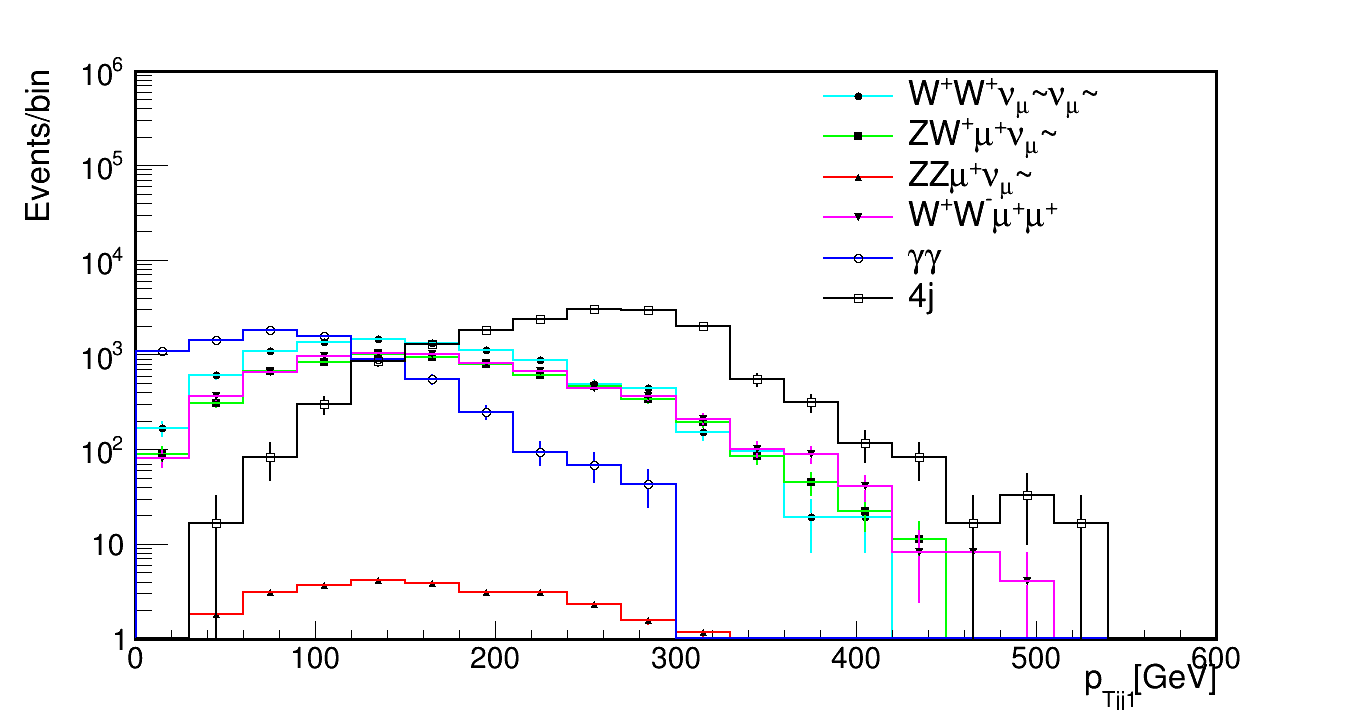}}
 \caption{Simulation results of hadronic channel, $\sqrt{s}=1\PTeV$, $\mathcal{L}=1\abinv$. (a) $M_{4j}$ distribution, (b) $M_{jj1}$ distribution, and (c) $p_{\mathrm{T},jj1}$ distribution.}
\label{fig5: Hadronic process variables distribution}
 \end{figure}

 After the determination of the preferred combination of the jet pairs, we compare the signals with backgrounds using the variables related to the two final reconstructed $\PW^{+}$ boson candidates to find criteria for further optimization. Fig.~\ref{fig5: Hadronic process variables distribution} shows the distributions of several selected variables: four jets invariant mass $M_{4j}$, the transverse momentum of one reconstructed boson, $p_{\mathrm{T},jj}$ and its mass, $M_{jj}$. The distributions show that $M_{4j}$ is the most important variable to distinguish between the signals and backgrounds, and there's significant discrepancy between the signals and backgrounds are also shown in the distributions of the other two variables. The summary of cuts in hadronic processes is given in Table~\ref{tab:my_table3}. The simulation results show that this channel is the most sensitive among the three channels at the same collision energy and luminosity.

\begin{table}[h]
     \centering
     \caption{The cut-flow table in the hadronic channel. }
     \label{tab:my_table3}
     \begin{tabular}{p{2.5cm}<{\centering}p{4cm}<{\centering}}
     \hline
       Variables &  limits  \\
        \hline
        $M_{4j}$ & $>600.0\PGeV$\\
        $p_{\mathrm{T},j_{1,2,3,4}}$ & $>50.0\PGeV$\\
        $p_{\mathrm{T},jj1}$ & $[20\PGeV,400\PGeV]$\\
        $p_{\mathrm{T},jj2}$ & $[50\PGeV,420\PGeV]$\\
        $M_{jj1}$ & $[50\PGeV,110\PGeV]$\\
        $M_{jj2}$ & $[10\PGeV,110\PGeV]$\\
        $|\eta_{j}|$ & $<4.7$\\      
        \hline
     \end{tabular}
 \end{table}

For the implementation of the Boosted Decision Tree (BDT) method, we shuffle the signal and background events in hadronic state, and define the training and test sets with the event ratio of 1:1. We apply the per-event weight $n_{\mathcal{L}X}=\sigma_{x}\mathcal{L}/N_{GX}$ during the training to account for the cross-section difference among the processes, where $\sigma_{x}$ is the cross-section of one process, $\mathcal{L}$ is the default target luminosity in this study (1\abinv\ ) for a 1\PTeV\, muon collider, and $N_{GX}$ is the total generated number of events~\cite{Yang:2021zak}. We use variables $M_{4j}, M_{jj1}, p_{\mathrm{T},jj1}, M_{jj2}, p_{\mathrm{T},jj2}$ as input features, namely reconstructed kinematics of each event are used for training.

\begin{figure}[ht!]
    \centering
    \includegraphics[width=7cm,height=5cm]{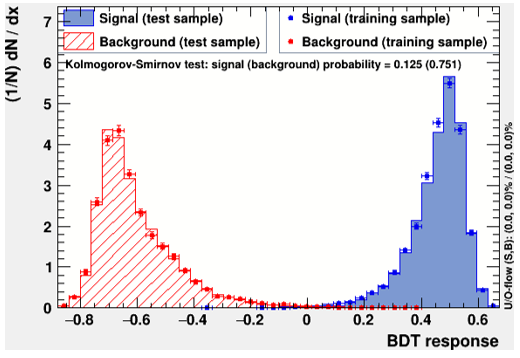}
    \caption{BDT training results in the hadronic channel at a $\sqrt{s}=1\PTeV$ muon collider. The Kolmogorov-Smirnov test results and score distributions for signal and background in the training and test are presented. }
    \label{fig:figure6}
\end{figure}

Fig.~\ref{fig:figure6} shows the results of BDT. We provide p-values from the Kolmogorov-Smirnov test and the BDT score
distributions for the signal and background in the training and test sets, as proof of no over-training in the BDT model. The receiver operating characteristic (ROC) curve of the trained model is then studied from the test sample, we find the background rejection is equal to 1, because the signal can be separated from the background completely when BDT score $>0.$ 

\subsection{Hadronic merged channel}

\begin{figure}[ht!]
    \centering
    \subfloat[\label{fig:a}]{
    \includegraphics[width=7cm,height=4cm]{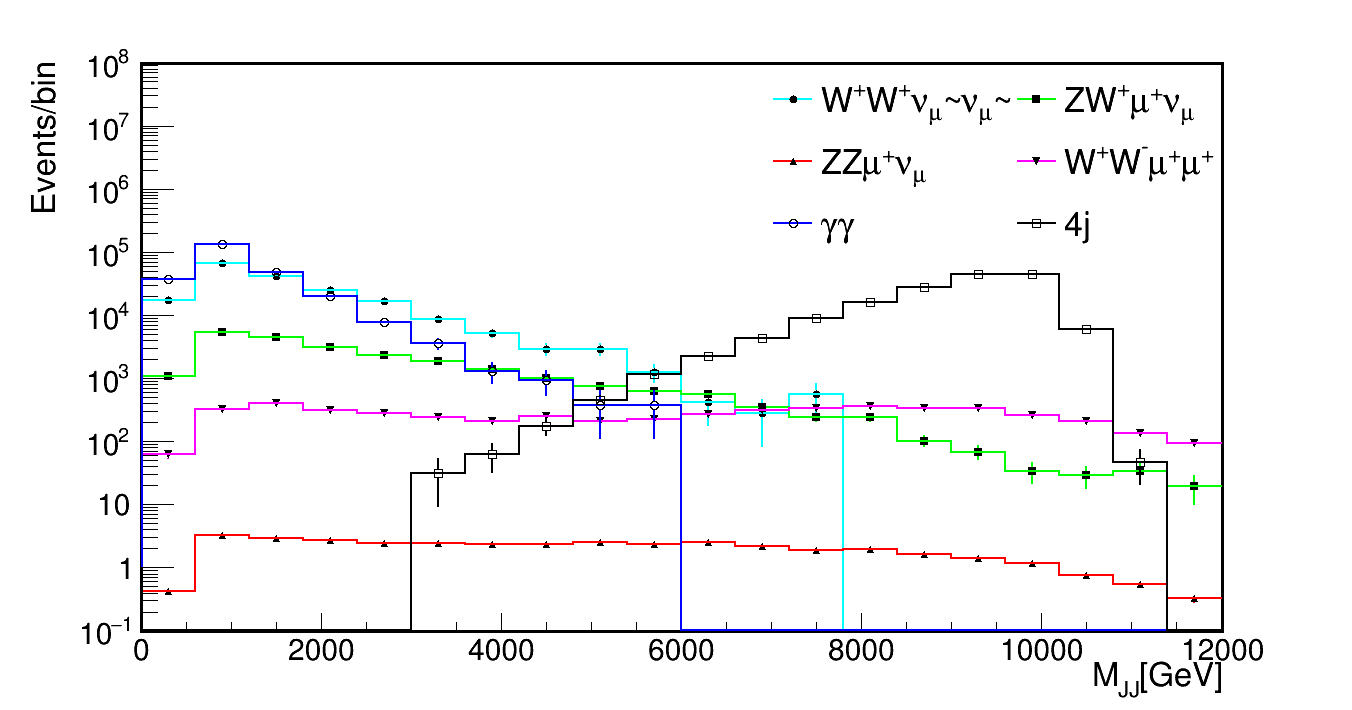}}
    \subfloat[\label{fig:b}]{
    \includegraphics[width=7cm,height=4cm]{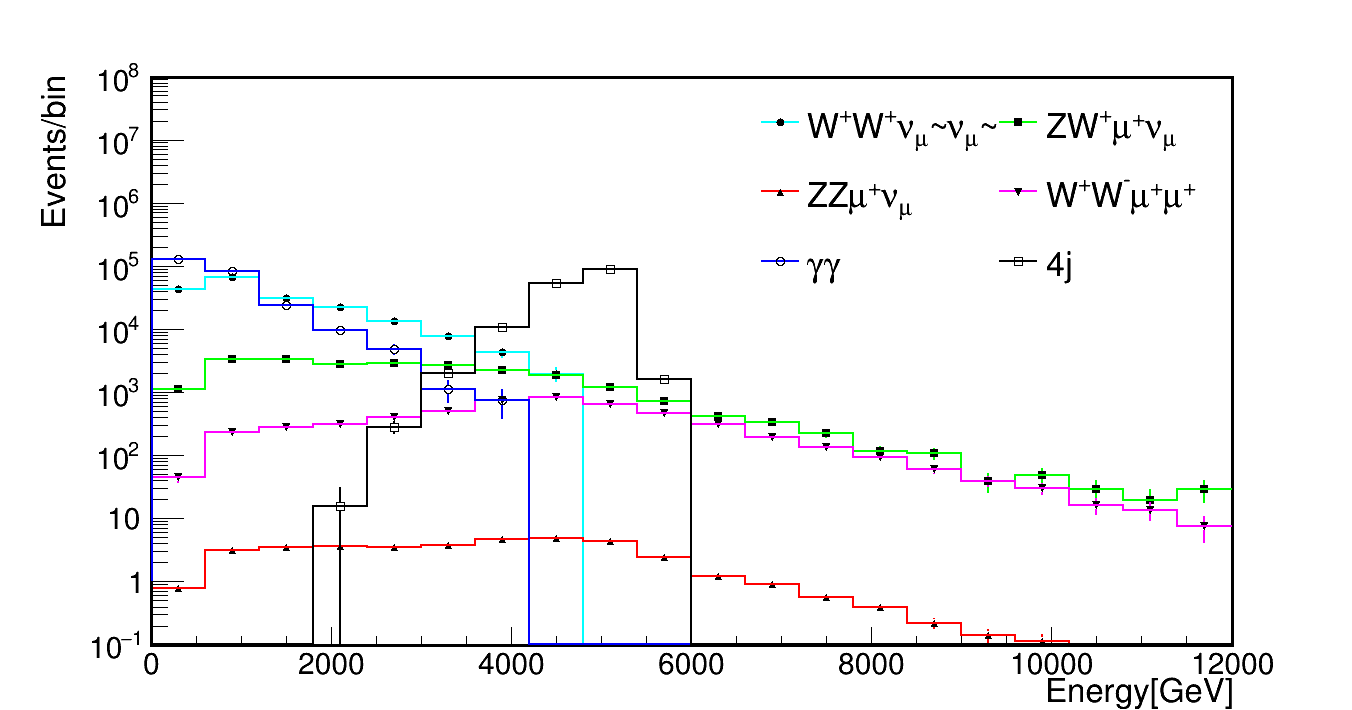}}
    \\
    \subfloat[\label{fig:c}]{
    \includegraphics[width=7cm]{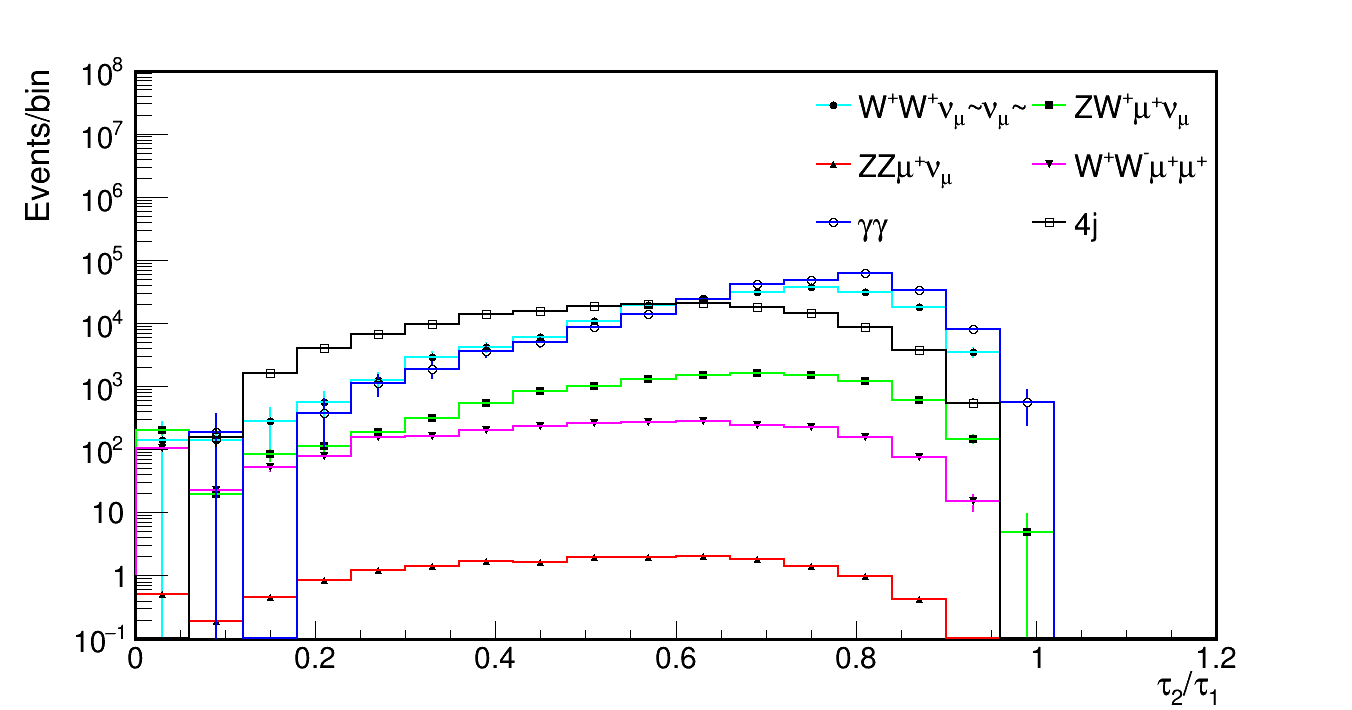}}
    \caption{Simulation results of hadronic channel, $\sqrt{s}=10\PTeV, \mathcal{L}=1\abinv$.
(a) the invariant mass of two fatjets, $M_{JJ}$ distribution, (b) the energy distribution of fatjet in SoftDropped algorithm, and (c) $\tau_{2}/\tau_{1}$ distribution.}
    \label{fig:my_figure7}
\end{figure}

When collision energy is several \PTeV, we must consider the boost effect when the two quarks from $\PW^{+}$ decay merged into a single fatjet with a mass around $M_{\PW}$. We use the processes at c.m. energy $\sqrt{s}=10\PTeV$ to research this scenario. Fig.~\ref{fig:my_figure7} shows some variables distribution of fatjet: the variables ``Energy" and ``$E_{t}$" are selected from SoftDroppedJet algorithm of fatjet, the variable $\mathrm{\tau_{N}}$ is named N-subjettiness, which is a parameter to veto additional jet emissions and define an exclusive jet cross-section. It can be calculated through the equation:

\begin{equation}
\tau_{N}=\frac{1}{d_{0}}\sum_{k} p_{\mathrm{T},k} \min{\left\{\Delta R_{1,k},\Delta R_{2,k}, \Delta R_{N,k} \right\}},\  d_{0}=\sum_{k} P_{T,k}R_{0}.   
\end{equation}

The $k$ runs over all constituent particles in a given jet, $p_{\mathrm{T},k}$ is the transverse momentum of one particle, $\Delta R_{J,k}$ is the distance in the rapidity-azimuth plane between a candidate subjet J and a constituent particle k. $d_{0}$ is the normalization factor, $R_{0}$ is the characteristic jet radius used in the original jet clustering algorithm. N-subjettiness can be used to effectively ``count" the number of subjets in a given jets~\cite{Thaler:2010tr}. In our analysis, we use $\tau_{2}/\tau_{1}$, which is a variable which can identify two-prong objects like boosted \PW\ boson, \PZ\ boson, and Higgs boson effectively. Table~\ref{tab:mytable4} gives the summary of cuts in this channel.

\begin{table}[h]
     \centering
     \caption{The cut-flow table in the hadronic channel, with $\sqrt{s}=10\PTeV.$ }
     \label{tab:mytable4}
     \begin{tabular}{p{3cm}<{\centering}p{6cm}<{\centering}}
     \hline
       Variables &  limits  \\
       \hline
        $\mathrm{Energy}$ & $[2000.0\PGeV,6000.0\PGeV]$\\
        $E_{t}$ & $[100\PGeV,6000\PGeV]$\\
        $\pt$ & $[100\PGeV,6000\PGeV]$\\
        $\mathrm{\tau_{2}/\tau_{1}}$ & $[0.05,0.95]$\\
        $M_{JJ}$ & $[3000\PGeV,11000\PGeV]$\\
        $|\eta_{FatJet}|$ & $<5.0$\\      
        \hline
     \end{tabular}
 \end{table}

\section{Results and Discussions}
\label{sec:results}
\begin{figure}
\centering
\includegraphics[width=15cm]{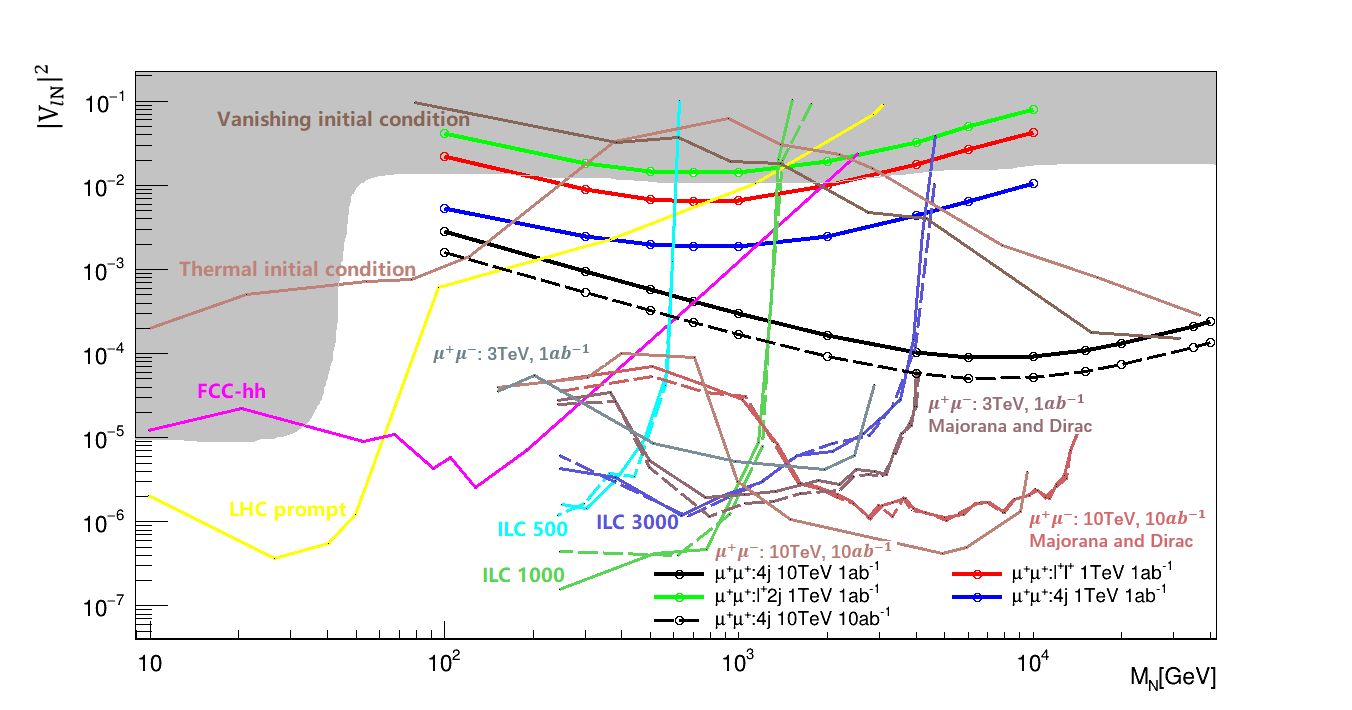}
\caption{2$\sigma$ exclusion limit of $|V_{\mu N_{1}}|^{2}$ as a function of varying Majorana neutrino mass $M_{N}.$ The green solid line corresponds to the semi-leptonic processes at a muon collider with $\sqrt{s}=1\PTeV, \mathcal{L}=1\abinv$. The red solid line corresponds to the pure-leptonic processes at a muon collider with $\sqrt{s}=1\PTeV, \mathcal{L}=1\abinv$. The dark-blue line corresponds to the hadronic processes at a muon collider with $\sqrt{s}=1\PTeV, \mathcal{L}=1\abinv.$ The black line corresponds to the hadronic processes at a muon collider with $\sqrt{s}=10\PTeV, \mathcal{L}=1\abinv.$ The black dotted line corresponds to the hadronic processes at a muon collider with $\sqrt{s}=10\PTeV, \mathcal{L}=10\abinv.$ The experimental result from LHC and other simulation results are also added for comparison.}
\label{fig:my_figure8}
\end{figure}

After analyzing four processes in three final states channels, we can obtain the exclusion limit of $|V_{\mu N_{1}}|$ of our signal channel: $\mu^{+}\mu^{+}\rightarrow \PW^{+}\PW^{+}.$ In Fig.~\ref{fig:my_figure8}, we present our simulation results of limit lines, as well as results from  other experiments and simulations. The red solid line corresponds to the pure-leptonic processes at a muon collider($\mu^{+}\mu^{+}$) with $\sqrt{s}=1\PTeV, \mathcal{L}=1\abinv$. The dark-blue line corresponds to the hadronic processes at a muon collider with $\sqrt{s}=1\PTeV, \mathcal{L}=1\abinv.$ The black line corresponds to the hadronic processes at a muon collider with $\sqrt{s}=10\PTeV, \mathcal{L}=1\abinv.$ The black dotted line corresponds to the hadronic processes at a muon collider with $\sqrt{s}=10\PTeV, \mathcal{L}=10\abinv.$ The solid brown and light-pink lines correspond to limits from considerations of viable leptogenesis scenarios~\cite{Drewes:2021nqr}. The grey area is the region excluded by a global scan~\cite{Chrzaszcz:2019inj}. The yellow line corresponds to the experimental limits from prompt trilepton searches at the LHC~\cite{Izaguirre:2015pga}. The pink line shows the simulated limits from a future FCC-hh~\cite{Antusch:2016ejd}. Three lines denoted ILC are simulated exclusion limits in future $e^{+}e^{-}$ linear colliders~\cite{Mekala:2022cmm}. Two groups of simulated results in $\mu^{+}\mu^{-}$ collider are also added~\cite{Kwok:2023dck,Li:2023tbx}. Our research gives simulated results for Majorana neutrinos masses range from 100 \PGeV\, to 40 \PTeV, it shows that better limitation is expected in the massive mass region ($M_{N}>10\PTeV$), especially with hadronic processes at $\sqrt{s}=10\PTeV.$

\section{Conclusions and Outlook}
\label{sec:conclusion}

In this paper, we investigate the potential of searching for Majorana neutrinos at future muon collider through the $\mu^{+}\mu^{+}\rightarrow \PW^{+}\PW^{+}$ scattering process. It is a typical $0\nu\beta\beta$ like process and can be used to research LNV phenomenon. In our simulation, we focus on the collider phenomenology of $\mu^{+}\mu^{+}\rightarrow \PW^{+}\PW^{+}$ process, to find the kinematic features that help to increase the detection potential, such as the distribution of $\cos{\theta_{\ell \ell}}$ in pure-leptonic processes and $M_{4j}$ in hadronic processes. We have studied three final states and four different conditions with fast simulation, determined the value of mixing elements squared $|V_{\mu N_{1}}|^{2}$ corresponding to various Majorana neutrino mass at CL=95\%. We also performed full simulation in pure-leptonic channel, the results show roughly similar distributions as fast simulation. Furthermore, we use BDT training on hadronic processes at $\sqrt{s}=1\PTeV$, the result shows that the variable $M_{4j}$ can be used to distinguish signal and backgrounds effectively. The distribution of some significant variables and associated cut-flow tables in pure-leptonic, semi-leptonic and hadronic channels are presented with collision energy $\sqrt{s}=1\PTeV$ and $\mathcal{L}=1\abinv$. We studied the fatjet signature at collision energy $\sqrt{s}=10\PTeV$, $\mathcal{L}=1\abinv$ and $10\abinv$ respectively, it turns out that these channels provide the strongest limitation. Compared with other researches as shown in Fig.~\ref{fig:my_figure8}, our analysis shows a unique advantage of using the same sign muon collider in searching of Majorana neutrinos, especially in the mass region above 10\PTeV. 

% \paragraph{Up to paragraphs.} We find that having more levels usually reduces the clarity of the article. 

%%%%%%%%%%%%%%%%%%%%%%%%%%%%%%%%%%%%%%%%%%%%%%%%%%%%%%%%%%%%

\begin{acknowledgments}
This work is supported in part by the National Natural Science Foundation of China under Grants No. 12150005, No. 12075004, and No. 12061141002, by MOST under grant No. 2018YFA0403900.
\end{acknowledgments}

%%%%%%%%%%%%%%%%%%%%%%%%%%%%%%%%%%%%%%%%%%%%%%%%%%%%%%%%%%%%
\appendix
\label{sec:appendix}
%%%%%%%%%%%%%%%%%%%%%%%%%%%%%%%%%%%%%%%%%%%%%%%%%%%%%%%%%%%%

% This is the most common positions for acknowledgments. A macro is
% available to maintain the same layout and spelling of the heading.

% \paragraph{Note added.} This is also a good position for notes added
% after the paper has been written.

% Bibliography

%% [A] Recommended: using JHEP.bst file
%% \bibliographystyle{JHEP}
%% \bibliography{biblio.bib}

%% or
%% [B] Manual formatting (see below)
%% (i) We suggest to always provide author, title and journal data or doi:
%% in short all the informations that clearly identify a document.
%% (ii) please avoid comments such as "For a review'', "For some examples",
%% "and references therein" or move them in the text. In general, please leave only references in the bibliography and move all
%% accessory text in footnotes.
%% (iii) Also, please have only one work for each \bibitem.

\end{document}